*Title*
MobiScout: A Scalable Cloud-Based Driving and Activity Monitoring Platform Featuring an IOS App and a WatchOS Extension.

*Names of authors / main developers (incl. affiliations, addresses, email)*
Kojo Konadu Adu-Gyamfi
Institute for Transportation, Iowa State University, 2711 s loop dr, Ames, IA 50010, USA. (kgyamfi@iastate.edu)

Karo Ahmadi-Dehrashid
Iowa State University, Ames, IA 50010, USA. (karo@iastate.edu)

Yaw Okyere Adu-Gyamfi
University of Missouri-Columbia, USA. (adugyamfiy@missouri.edu)

Pujitha Gunaratne
Toyota research, USA. (pujitha.gunaratne@toyota.com)

Anuj Sharma
Institute for Transportation, Iowa State University, 2711 s loop dr, Ames, IA 50010, USA. (anujs@iastate.edu)

**Abstract**
MobiScout is an iOS software that monitors users' driving habits and physiological conditions while on the road. The Mobiscout app was created to provide a low-cost next-generation data collection and analysis solution for naturalistic driving studies. MobiScout collects real-time data, including physiological information from drivers in their normal driving conditions using sensors and cameras on mobile phones, smartwatches, and Bluetooth-enabled OBD equipment. The MobiScout software captures vehicle and driving data, including speed, braking, pulse rate, and acceleration, while the phone's camera captures everything inside and outside the car. Data captured can be streamed to cloud storage in real-time or persisted in local storage in WIFI dead zones. The information gathered will be studied further to better understand typical traffic behavior, performance, surroundings, and driving context among drivers.

**Keywords**
Swift; IOS; OBD (Onboard Diagnostics); Naturalistic Driving; Storyboard; UI (User Interface)

## 1. Motivation and significance

Naturalistic driving studies (NDS) record data from the driver, vehicle, and environment to gain insight into day-to-day driving behaviors [7]. The usage of data acquisition systems (DAS) to continuously monitor driving behaviors is used to collect data for NDS [6]. For naturalistic driving research, these data collecting methods provide expensive in-vehicle sensors and technology. However, instead of an in-vehicle DAS, mobile devices provide a low-cost driver data collecting solution [6]. Naturalistic driving studies can be scaled to a larger number of people at a lesser cost using smartphones, advancing NDS in road safety. Data acquisition for naturalistic driving studies can be used with additional data such as physiological conditions as new smartphone sensors are introduced, and also managed for detecting events and conditions between normal driving behavior and crash or near-crash events using the MobiScout mobile application. As a result, MobiScout data can be utilized to provide information on the types of road behaviors and the causes for these behaviors. Users can engage with the MobiScout mobile app through a user-friendly user interface (UI). The user must first register and agree to the terms and conditions to use the mobile application. The user can start the app feature by pressing the record start button on the activity monitoring page and setting the phone facing them on the dashboard of their automobile for effective and efficient in-vehicle and road video recording of the driver and its environment according to Mobiscout. With the availability of WIFI, data is automatically streamed to the cloud, while data is stored locally in WIFI-dead zones.

## 2. Software description

MobiScout mobile app is developed for smartphones with the capability of recording the output from front and back cameras simultaneously and using iOS as an operating system. MobiScout is written using swift programming language and XCode as the integrated development environment. Swift is used since it is powerful and intuitive, safe and concise by design. Apple development documents and its guides are used as references during the programming of Mobiscout. The local database used for local data persistence is AWS Amplify Datastore. Amplify Datastore is used because it simplifies working with distributed and cross-user data
.
The app's operation is carried out by a user (driver) who accesses MobiScout after successful registration into the app. Mobiscout is used by the user who, during driving, uses the application to record vehicle and driving data, including driver and road view video, sensor, health, and vehicle CAN-BUS data. Fig. 1. Provides a flowchart for the operation of MobiScout. Rectangular boxes indicate the screens of the application.

### 2.1. Software architecture:

The iOS design pattern and coding structure were used to construct the MobiScout mobile application, which was written in swift. Model View Controller (MVC), an architectural design paradigm advocated by Apple [1], is used to create MobiScout. The Model View Controller (MVC) design pattern dictates that an application must contain a data model, presentation information, and control information [2]. To adhere to MVC, these entities must be split into separate objects. The model objects, view objects, and controller objects that make up the MVC architecture shown in Fig. 2. are the three major design elements that will be broken down into their implementations in the software architecture section.

#### 2.1.1. Model:
The model object contains MobiScout's data and defines the logic which manipulates that data. This logic is encapsulated in model objects. The data model object for MobiScout consists of a health data model, vehicle data model, and sensor data model. The health data model presented in Fig. 3**Error! Reference source not found.** represents the user's personal health information (PHI). This model consists of the average heart rate, and total distance walked, step count, audio-visual exposure. These health data are queried from iOS HealthKit [3] repository. Data is sampled five days before the data collection to provide an overview of the user's physiological condition before driving. The second model object implemented is the vehicle data model. This model consists of the data schema for the user's vehicle information obtained from the VIN. Finally, as shown in Fig. 5, the sensor data model Fig. *5*consists of time series data collected from various sensors of the users' phones and vehicles. This model data originates from the smartphone onboard motion and location (GPS) sensors and vehicles onboard diagnostics data.

### 2.1.2. View:

The view object model represents the MobiScout user interface. The view model displays the model's data to the user. The view is aware of how to access the model's data but is unaware of what that data means or what the user might do with it. The XCode storyboard functionality was used to create the MobiScout user interface. With the help of the XCode tool storyboarding, you can visually create each of the various iOS app displays and the flow of information between them. MobiScout user interface as shown in Fig. 12. includes user authentication, activity monitoring, library, upload, chart, and settings screens.

The user authentication screen, as presented in Fig. 6. provides the system which manages the user registration and login processes. Because Mobiscout gathers sensitive information from its users, it is necessary to offer a secure mobile application that will authenticate any users wanting to use MobiScout's features.

The next screen after user authentication is MobiScout's main landing menu upon successful login into the application. The main landing menu is designed using Apple's iOS UITabBarController [4]. The tab bar controller is a view that lets you switch between tab bar elements within the same view using a multiselection interface. This view style was employed to make it simple to navigate MobiScout and show a user-friendly user interface. Driver and road activity monitoring, library, upload, charts, and settings views are available through the tab bar menu.

**Driver and Road Activity Monitoring Screen:** This screen in Fig. 7. shows back and front camera video previews. Labels on the screen also display the user's heart rate, vehicle speed, acceleration z-direction, and OBD connection status. It features a record button that starts the app.

**Library Screen:** The library screen, as shown in Fig. 8. is implemented using Apple's iOS UITableViewCell. The table view cells show table rows with cells that visually have selectable items. In the library, you can view details like the date the gathered multi-camera footage was created and other pertinent data. The user interface options on this panel allow you to upload, preview recorded videos, and open data files. Additionally, each cell of the table view contains these options.

**Upload Screen:** The Upload screen of Mobiscout, as shown in Fig. 9. has a progress bar in the user interface, labels for the file's name, and the amount of data that has been uploaded. On the upload page, UI buttons allow you to pause or stop data transfers. Each upload is kept in a cell of the table view on iOS.

**Chart Screen:** In the chart view shown in Fig. 10., a line chart based on data from the Core Motion sensor depicts users' driving behaviors.

**Settings Screen:** Each settings choice is represented in the settings screen as shown in Fig. 11. by UI labels, sliders, and buttons.

### 2.1.3. Controller:

The controller uses every component in the model layer to define the information flow in the MobiScout app. The controller serves as the interface between the application's model and display. It pays attention to events that the view triggers and reacts appropriately. View controllers are used in the Mobiscout app for each screen. The controllers of Mobiscout include the following:

**Login view controller:** The login view controller is in charge of a login view comprising several sub-views. The login view will allow the user to log in to the Mobiscout application. The user's email address and password are entered into text boxes on the login screen. It also contains a Mobiscout registration button and a Google social sign-in button. This login page is set up to manage users and their properties using Amazon Amplify Cognito User Pools [8]. A user interface was designed to get the user's username and password. After the user provides their username and password, the sign-in flow begins by activating authentication methods from the AWS Amplify library [8].

**Registration view controller:** The MobiScout mobile application's user registration page allows users to sign up for it. A valid email address and password must be provided to register a user. The registration process is launched using the AWS Amplify API [9]. The next step in the registration procedure is the user's confirmation. An email with a confirmation code will be sent to the address entered during registration. After successfully registering, the user will be prompted to input the confirmation code they have received via email in the text area of the confirmation screen.

**Confirmation view controller:** The task of the confirmation controller is to supervise the confirmation process. The next step in the registration procedure is the user's confirmation. The email address provided upon registering will receive an email containing a confirmation code. Enter the confirmation code you received through email in the confirm Signup call on the confirmation screen view.

**Main view controller:** The driver monitor and road view activity recording screen will be the default landing screen after a user logs in or registers to use the MobiScout mobile app for the first time. The record button on this screen allows the user to launch the MobiScout application and view their driving data. This screen shows video previews from the front and back cameras. Additionally, it contains labels that display the driver's heart rate, vehicle speed, and acceleration in the z-direction. The front and back cameras' output were simultaneously recorded into video files utilizing a multi-camera capture session of the IOS AVCaptureMultiCamSession [10]. Only iPad Pro with an A12X CPU and iPhones with an A12 CPU or later can use this feature.

**Library view controller:** The library screen displays information such as the date the acquired multi-camera video was created and other related data. From the library interface, the user can explore collected movies and data files, upload files and data files, and remove videos and data files. A table view controller composed of table view cells is used to do this. Table views in iOS display vertically scrolling content in rows in a single column. Each row in the table contains one user element who has saved data files for each time and date. There are buttons in each row for opening, uploading, and deleting data files.

**Upload view controller:** The upload screen shows the progress of users' data uploads. The upload interface will allow the user to postpone or cancel data uploads. The upload screen will display the user's current data uploads, upload history, and progress. This is performed by leveraging IOS's table view to building each cell for each upload. Each upload cell has a pause and cancel button. Pause internally uses the NSURLSessionTask suspend API, which only temporarily suspends the process and does not completely halt the transfer. The upload operation calls the cancel function to cancel the upload (for example, when the user presses the Cancel button).

**Graph view controller:** Charts depicting users' driving actions based on their Core Motion sensor data can be found on the graph screen (including for accelerometer, gyroscope, pedometer, and environment-related events). This is done with the help of the charts pod. The graph screen's underpinning view is set to the line chart view class.

**Settings view controller:** To alter user settings, the view controller uses the iOS user defaults methods. For each of the options, the iOS UI view was used. Each cell has its own set of buttons and labels. The changes are saved by clicking the 'set new value' button.

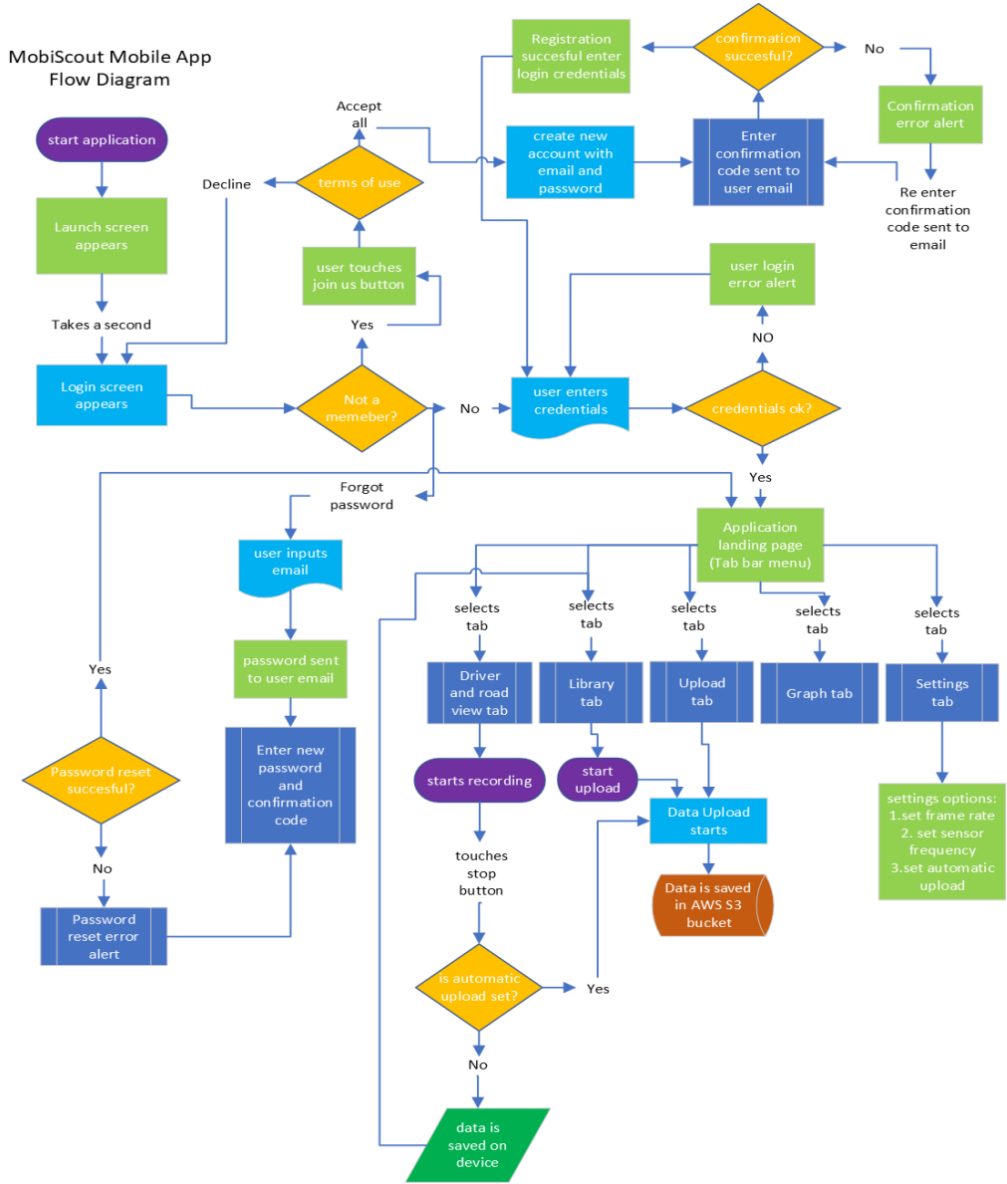

Fig. 1. Flow Chart MobiScout App Operation

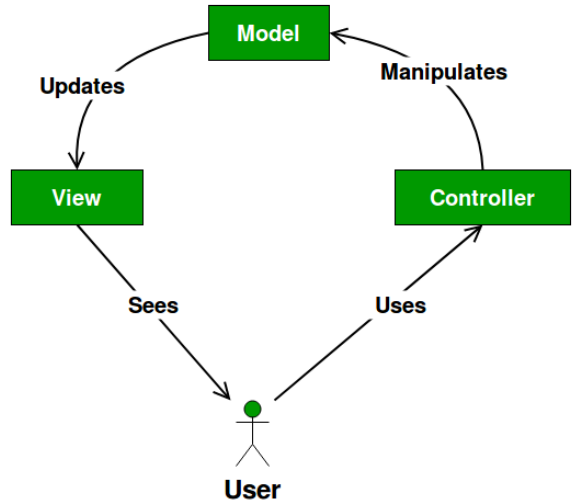

Fig. 2. MVC Design (source: https://www.geeksforgeeks.org/mvc-design-pattern/)

```
class HealthData: Codable{
    var HeartRate:[Double] = []
    var HeadphoneAudioExposure:[Double] = []
    var DistanceWalkingRunning:[Double] = []
    var StepCount:[Double] = []
}
```

Fig. 3. Health data model object

```
class UserDefaultsData : Codable{
    //user frame rate
    var frameRate : Double!
    //user frequency
    var frequency : Double = 1
    //upload data automatically after recording
    var automaticUpload : Bool = true
}
```

Fig. 4. User defaults data model

```
class SensorData: Codable{

    //MARK: Add device motion data here

    //latitude
    var latitude : Double!
    //longitude
    var longitude : Double!
    //accelerometer data x
    var accelerationX : Double!
    //accelerometer data y
    var accelerationY : Double!
    //accelerometer data z
    var accelerationZ : Double!
    //gyro data x
    var gyroDataX : Double!
    //gyro data z
    var gyroDataY : Double!
    //gyro data z
    var gyroDataZ : Double!
    //pitch
    var pitchData : Double!
    //roll
    var rollData : Double!
    //yaw
    var yawData : Double!
    //quadrant Data x
    var quaternionX : Double!
    //quadrant Data y
    var quaternionY : Double!
    //quadrant Data z
    var quaternionZ : Double!
    //quadrant Data w
    var quaternionW : Double!
    //gravity data
    var gravityDataX : Double!
    //gravity data
    var gravityDataY : Double!
    //gravity data
    var gravityDataZ : Double!
```

Fig. 5. Sensor data object model

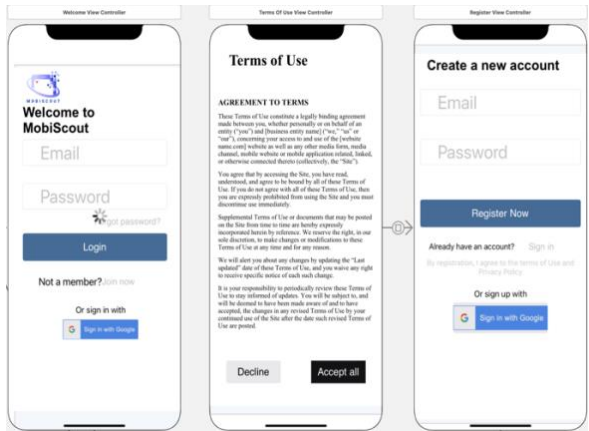
Fig. 6. User authentication Screen

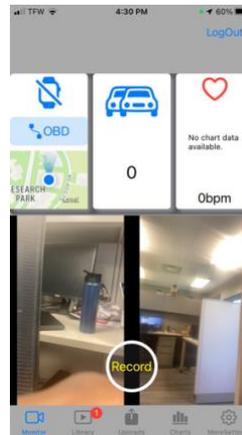
Fig. 7. Driver monitor and road activity Screen

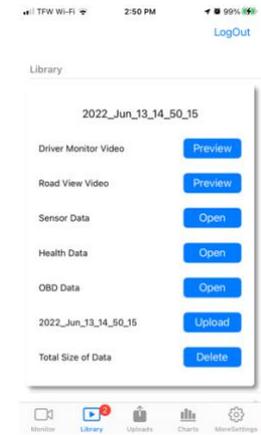
Fig. 8. Library Screen

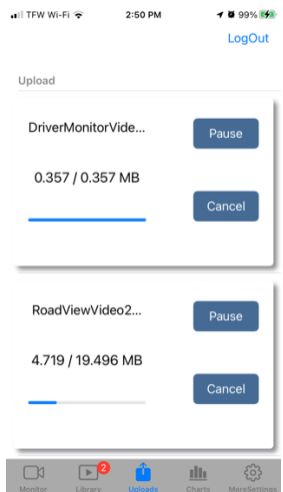
Fig. 9. Upload Screen

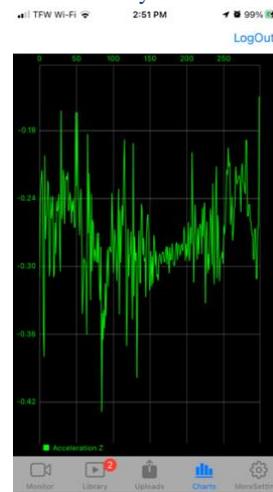
Fig. 10. Chart Screen

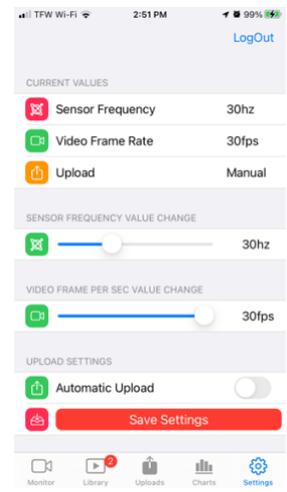
Fig. 11. Settings Screen

Fig. 12. Mobiscout User Interface

### 2.2. Software functionalities:

MobiScout is used to provide a low-cost platform for tracking users' driving activities and physiological states while on the road in order to conduct naturalistic driving studies. Drivers will utilize the smartphone app to track their physiology, driving activities, and health data in real-time by interacting with external devices such as a smartwatch and a Bluetooth-enabled OBD device. Each user's information will be uploaded to a cloud-based database. MobiScout's cloud architecture will enable advanced analytics and, as a result, deliver meaningful insights from the obtained data. This system will provide a secure framework for gathering and evaluating naturalistic driving data that is physiology-aware.

The major functionalities of the mobile application are as follows:

- Provide users with compatible mobile and wearable devices with a comprehensive, user-friendly platform for recording in-vehicle driver monitor and road view video, location and motion data, and health data.
- Users are registered using a simple method, and further useful information is obtained later.
- Allow users to grant authorization for data collection.
- When a user's attempted registration is still incomplete, send them an app notice.
- Minimize the distraction caused by the app while driving.
- Provide an intuitive user experience for data collection.
- Revalidate data in a test file before uploading it to the cloud. Validate each user's data submitted to the cloud following their privacy settings and notify any data submission problems.
- Allow for the automation of uploads and smooth integration with previously stored data in the cloud.
- Receive and verify data that the app publishes over the platform that the user has corrected.
- Receive data from users in several ways and at different frequencies, such as real-time transmission via mobile services and manual file upload within the mobile app.
- Accept any data that complies with the established data standards.
- To ensure the secure handling of sensitive and protected data, encrypt all data.
- Standardize data to improve quality and consistency, for example, by using standard units and statistical measures.
- Be a versatile platform that can grow to suit the demands of future mobile applications.

## 3. Illustrative examples

To test how the MobiScout mobile application works as intended, the mobile app was tested by the research team. Mobiscout was tested using iPhone 12, apple smart watch, and Bluetooth-enabled OBD dongle. During the test, the mobile device, as shown in Fig. 16. was mounted on the dashboard of the user's vehicle, the user had an Apple smart watch series six, as shown in Fig. 14. and a Bluetooth-enabled dongle presented in Fig. 15 was also inserted into the OBD port of the user's vehicle. Detailed steps for the illustration of the Mobiscout mobile app are as follows:

1. **User Registration and login:** The tester registers for MobiScout by providing an active email address and a password. Before registering, the tester must agree to the terms of service (including the privacy policy) and be presented with a Terms of Service screen. The Terms & Conditions of the Mobiscout app explain what is expected of both the app and its users. Users have the option to accept or decline the offer. The next step in the registration procedure is the user's confirmation. A confirmation code will be sent to the email address entered during registration. The user inputs the confirmation code received through email in the text box of the confirmation page that will be given to the user after a successful registration during the confirm signup call.

2. **User privacy:** The landing page will be the main display after users log into the MobiScout mobile app. This primary display is used to interact with the MobiScout app's features. Because health and user data can be sensitive, iOS gives users fine-grained control over the information that apps can disclose before allowing them to utilize the app's capabilities. Each MobiScout must be expressly granted permission to read data from the mobile device and smartwatch by the user. Users can grant or refuse permission for each category of data separately.

3. **App functionality:** The tester checked that the smartwatch had been correctly linked with the mobile device before launching the application. In addition, the OBD dongle was installed in the vehicle. The tester initiates MobiScout driver activity monitoring by pressing the record button on the driver and road activity monitoring tab bar item. The video previews of the front and back cameras are covered while recording to reduce driver distraction and maintain safety, as illustrated in Fig. 17.

    Regarding video data collection, the smartphone camera could simultaneously capture video at a frame rate of 30 frames per second for both front and back capture devices. This ensured high-quality video outputs, allowing quick and efficient data processing for object and event detection analyses. Furthermore, because the smartphone's inbuilt algorithms preprocess the raw sensor readings, motion sensors produce unbiased data collection. Additionally, motion data from the smartphone can be adjusted to a specific update frequency interval, ensuring that all motion-related events are captured regularly. The user's location data was also captured using the smartphone's basic location devices. These devices offered standard and major location updates with the highest possible accuracy using additional sensor data. While the required accuracy is 50m, it can reach as good as 5m precision. However, if Mobiscout is not authorized to access exact location services, location services may provide less accurate data. One of the platform's standout features was its ability to connect to a smartwatch and capture real-time heart rate data. Mobiscout read the user's heart rate at 5-second intervals. Data flow between Mobiscout and peripherals to cloud storage is presented in Fig. 13.

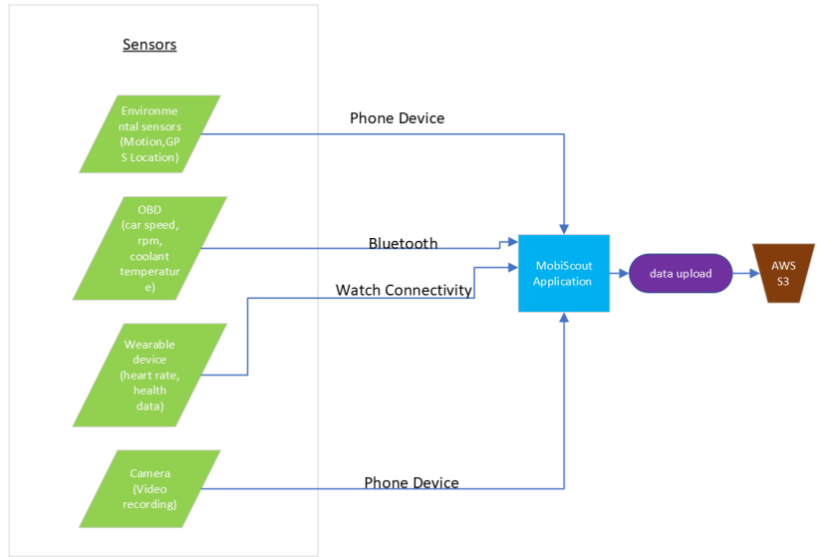

Fig. 13. Mobiscout Data Flow

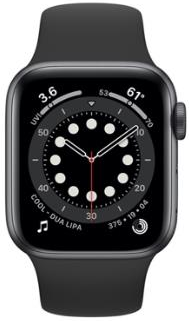

Fig. 14. Apple Watch(source: Apple Watch)

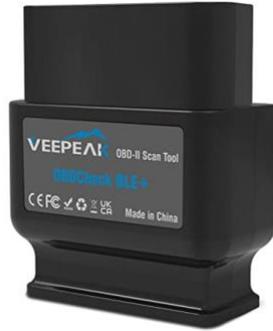

Fig. 15. OBD2 dongle (source: OBD2)

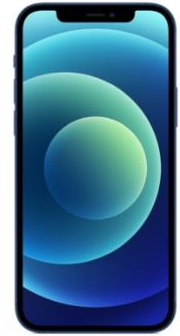

Fig. 16. Iphone12 (source: phone)

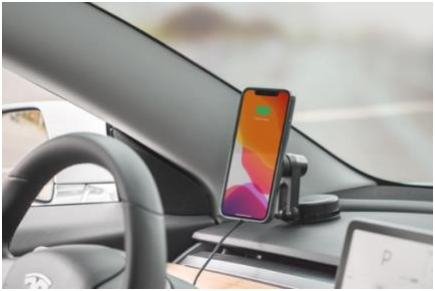

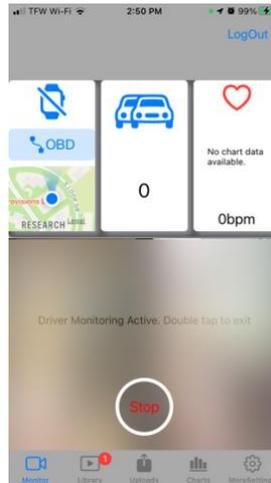

Fig. 17. Driver Activity Monitoring

## 4. Impact

The Mobiscout smartphone application provides access to a next-generation physiology-aware naturalistic driving platform and an activity data gathering and analysis platform. Existing naturalistic driving studies [5,7] that sought to provide insight into the everyday natural behavior of drivers can be pursued cost-effectively with the MobiScout mobile app, which leverages the onboard sensors of a smartphone device rather than the hardware-driven black boxes used in traditional naturalistic driving studies. Using MobiScout, we can extend data collecting to the next level, reaching tens of thousands of people worldwide. As the number of available sensors and smartphone capabilities increase, it will be simple to give improved functionality with the touch of a button. It should be highlighted that smartwatch data is captured beyond the time spent driving and can be used for preliminary driver evaluations. The data will be hosted in the cloud, and the data analytics capabilities will be scalable as data sizes rise.

## 5. Conclusions

MobiScout is a mobile application for collecting and analyzing activity and driving data that is physiology-aware. MobiScout is affordable and upgradable, scalable, and user-friendly.

Mobiscout has a user-friendly interface that is easy to traverse, thanks to a customizable tab bar menu. Mobiscout allows for low-cost data collecting for naturalistic driving studies, automated cloud data storage, and monitoring of the driver's physiological condition. MobiScout development is continuing to focus on implementing various in-app analytical tools, such as machine learning algorithms, to automatically deliver insights to data acquired on the device, reducing the amount of data transmitted to cloud storage by only uploading the information gathered on the application.